\documentclass[prd,aps,eqsecnum,floatfix,nofootinbib,preprint,tightenlines]{revtex4}

\usepackage{latexsym}
\usepackage{graphicx}
\usepackage{amsmath}
\usepackage{amssymb}

\usepackage[pdftex]{color}
\usepackage[colorlinks=true,linkcolor=blue,filecolor=blue,urlcolor=blue,citecolor=blue,pdftex=true,plainpages=false]{hyperref}

\def\bibi{\bibitem}

\let\inodot=\i

\def\a{\alpha}

\def\d{\delta}
\def\g{\gamma}

\def\i{\iota}

\def\k{\kappa}
\def\l{\lambda}
\def\m{\mu}
\def\n{\nu}

\def\p{\pi}                     % Also, \varpi
\def\P{\Pi}

\def\cbo{{\,\raise-.15ex\Sc [\,}}                       % curly "
\def\ltap{\raisebox{-.4ex}{\rlap{$\sim$}} \raisebox{.4ex}{$<$}}   % < or ~
\def\ddt#1{{\buildrel {\hbox{\LARGE .\kern-2pt.}} \over {#1}}}% double dot-over
\def\ie{\mbox{\it i.e.}}

\def\etc{\mbox{\it etc.}}

\def\tr{{\rm tr}\,}

\def\half{{1\over 2}}

\def\floatcaption#1#2{ \caption{ #2 \ [#1] \label{#1}} }
\def\floatcaption#1#2{ \caption{#2 \label{#1}} }

\def\ttl#1{{\it #1}}
\def\ttl#1{}

\long\def\symbolfootnote[#1]#2{\begingroup%
\def\thefootnote{\fnsymbol{footnote}}\footnote[#1]{#2}\endgroup}

\long \def \blockcomment #1\endcomment{}

\def\etc{{\it etc.}}
\def\seef{{\it cf.}}

\def\qbar{{\overline{q}}}

\def\hp{{\hat{p}}}
\def\hth{{\hat{\theta}}}
\def\hmu{{\hat{\mu}}}
\def\hnu{{\hat{\nu}}}

\begin{document}

\begin{center}
\begin{boldmath}
{\large\bf The hadronic vacuum polarization with\\ twisted boundary conditions}\\[0.4cm]
\end{boldmath}
\vspace{3ex}
{Christopher~Aubin,$^a$ Thomas~Blum,$^b$
Maarten~Golterman,$^c$%
\symbolfootnote[2]{Permanent address: Department of Physics and Astronomy, San Francisco State University, San Francisco, CA 94132, USA}  
Santiago~Peris$^d$
\\[0.1cm]
{\it
\null$^a$Department of Physics and Engineering Physics\\ Fordham University, Bronx,
NY 10458, USA\\
\null$^b$Physics Department\\
University of Connecticut, Storrs, CT 06269, USA\\
\null$^c$Institut de F\'\inodot sica d'Altes Energies (IFAE), 
Universitat Aut\`onoma de Barcelona\\ E-08193 Bellaterra, Barcelona, Spain\\
\null$^d$Department of Physics, Universitat Aut\`onoma de Barcelona\\
E-08193 Bellaterra, Barcelona, Spain}}
\\[6mm]
{ABSTRACT}
\\[2mm]
\end{center}
\begin{quotation} 
The leading-order hadronic contribution to the muon anomalous magnetic moment is given by a weighted integral over the subtracted hadronic vacuum polarization.
This integral is dominated by euclidean momenta of order the muon mass, \ie, momenta not accessible on current lattice volumes with periodic boundary conditions.   Twisted boundary conditions can in principle help in accessing
momenta of any size even in a finite volume, but their use leads to a modification of the Ward--Takahashi identity that normally guarantees transversality of the vacuum polarization.   As a result, the vacuum polarization
contains a non-transverse, quadratically divergent term, which arises as an
artifact of using twisted boundary conditions in a finite volume.   In this article,
we show how to determine and remove this term from the vacuum polarization.
\end{quotation}

\vfill
\eject
\setcounter{footnote}{0}

%%####%%
%\newpage
\section{\label{introduction} Introduction}
%%####%%
The leading-order hadronic (HLO) contribution to the anomalous magnetic moment $a_\m=(g-2)/2$ of the muon is given by the integral \cite{TB2003,ER}\footnote{For an overview of lattice computations of the muon anomalous
magnetic moment, see Ref.~\cite{TB2012} and references therein.}
\begin{eqnarray}
\label{amu}
a_\m^{\rm HLO}&=&4\a^2\int_0^\infty dp^2\,f(p^2)\left(\P^{\rm em}(0)-\P^{\rm em}(p^2)\right)\ ,\\
f(p^2)&=&m_\m^2 p^2 Z^3(p^2)\,\frac{1-p^2 Z(p^2)}{1+m_\m^2 p^2 Z^2(p^2)}\ ,\nonumber\\
Z(p^2)&=&\left(\sqrt{(p^2)^2+4m_\m^2 p^2}-p^2\right)/(2m_\m^2 p^2)\ ,\nonumber
\end{eqnarray}
where $m_\m$ is the muon mass, and for non-zero momenta
$\P^{\rm em}(p^2)$ is defined from the hadronic contribution to the
electromagnetic vacuum polarization $\P^{\rm em}_{\m\n}(p)$:
\begin{equation}
\label{Pem}
\P^{\rm em}_{\m\n}(p)=\left(p^2\d_{\m\n}-p_\m p_\n\right)\P^{\rm em}(p^2)
\end{equation}
in momentum space.   Here $p$ is the euclidean momentum flowing through the vacuum polarization.

The integrand in Eq.~(\ref{amu}) is dominated by momenta of order the muon
mass; it typically looks as shown in Fig.~\ref{f1}, with the peak located at
$p^2\approx (m_\m/2)^2$.   For a precision computation of this integral
using lattice QCD, one would therefore like to access the region of this peak.
In a finite volume with periodic boundary conditions, the smallest available non-vanishing
momentum is $2\p/L$, with $L$ the linear size of the lattice volume.
Setting $2\pi/L\approx m_\m/2$ leads to a value of $L$ equal to
about 25~fm, which is out of reach of present lattice computations, if the
lattice spacing $a$ is chosen to be such that one is reasonably close to the
continuum limit.
Clearly, a different method for reaching such small momenta is needed.
In this article, we discuss the use of twisted boundary conditions in order
to vary momenta arbitrarily in a finite volume.

Twisted boundary conditions have already been used in order to
access the connected part of $\P^{\rm em}(p^2)$ at momenta smaller than $2\p/L$ \cite{DJJW2012}.\footnote{In Ref.~\cite{DJJW2012},  $\Pi^{\rm em}(q^2)$  is extracted from the off-diagonal components of the vacuum polarization tensor, so the contact terms to be discussed here do not contribute to $\Pi^{\rm em}_{\mu\nu}$.}
However, as we will show here, any current used in the definition of
$\P^{\rm em}_{\m\n}(p)$ with twisted boundary conditions cannot be conserved,
and thus $\P^{\rm em}_{\m\n}(p)$ is necessarily not purely transverse.   In other words,
in the presence of twisted boundary conditions, $\P^{\rm em}_{\m\n}(p)$
cannot be written as in Eq.~(\ref{Pem}) above.   The relevant Ward--Takahashi identity (WTI) gets modified by twisting, leading to an extra term proportional
to $\d_{\m\n}$ in the vacuum polarization.   While this extra term is a finite-volume artifact, it turns out to be quadratically divergent, and thus a 
potentially significant obstruction to the extraction of $\P^{\rm em}(p^2)$
from $\P^{\rm em}_{\m\n}(p)$.   

This article is organized as follows.   In Sec.~\ref{twisted}, we briefly review the application of twisted boundary conditions to the computation of the vacuum polarization with arbitrary momentum.   Then, in Sec.~\ref{WI}, we formulate the
WTI, and demonstrate that this identity contains a contact term originating from the fact that any current used in order to define the vacuum polarization with non-zero twist is necessarily not conserved in a finite volume.   This leads to the appearance of non-transverse terms in the vacuum polarization, and in Sec.~\ref{subtraction} we show how these can be computed and subtracted, in order to allow the determination of $\P^{\rm em}(p^2)$.   In Sec.~\ref{numerical} we verify that indeed the WTI is satisfied numerically on a 
typical gauge configuration, and we have a first look at the numerical size
of the contact term relative to the complete vacuum polarization.
Section~\ref{conclusion} contains
our conclusions, and an appendix verifies the WTI to leading order in weak-coupling perturbation theory.

\begin{figure}[t]
\centering
\includegraphics[width=4in]{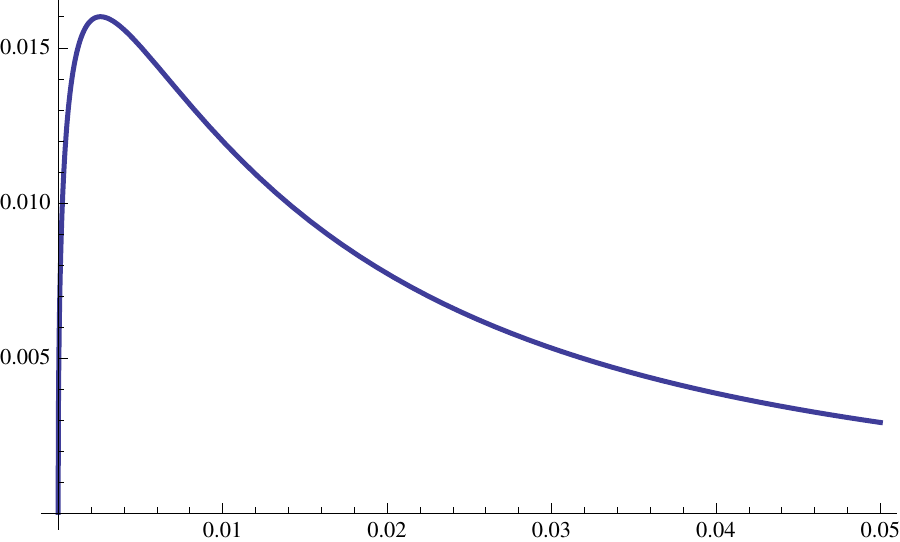}
\floatcaption{f1}{Typical shape of the integrand in Eq.~(\ref{amu}), with $p^2$ in
GeV$^2$ on the horizontal axis, and arbitrary units on the vertical axis.}
\vspace*{2ex}
\end{figure}
 
%%####%%
%\newpage
\section{\label{twisted} Twisted boundary conditions}
%%####%%
The aim is to compute the connected part of the two-point function of the 
electromagnetic current, 
\begin{equation}
\label{emcurrent}
J^{\rm em}_\m(x)=\sum_i Q_i \qbar_i(x)\g_\m q_i(x)\ ,
\end{equation}
in which $i$ runs over quark flavors, and quark $q_i$ has charge $Q_i e$,
in a finite volume, but with an arbitrary choice of momentum.   In order to do
this, we will employ quarks  which satisfy twisted boundary conditions
\cite{PB,GDRPNT,CSGV},
\begin{subequations}
\label{twistedbc}
\begin{eqnarray}
q_t(x)&=&e^{-i\theta_\m}\,q_t(x+\hmu L_\m)\ ,\label{twistedbca}\\
\qbar_t(x)&=&\qbar_t(x+\hmu L_\m)\,e^{i\theta_\m}\ ,\label{twistedbcb}
\end{eqnarray}
\end{subequations}
where the subscript $t$ indicates that the quark field $q_t$ obeys twisted boundary
conditions, $L_\m$ is the linear size of the volume in the $\mu$ direction
($\hmu$ denotes the unit vector in the $\mu$ direction), and
$\theta_\m\in [0,2\p)$ is the twist angle in that direction.\footnote{If anti-periodic boundary conditions are used in the time direction, that corresponds
to the choice $\theta_4=\pi$.}
   For a plane wave $u(p)e^{ipx}$,
boundary condition~(\ref{twistedbca}) leads to the allowed values
\begin{equation}
\label{allowed}
p_\m=\frac{2\p n_\m+\theta_\m}{L_\m}\ ,\qquad n_\m\in\{0,1,\dots,L_\m-1\}\ .
\end{equation}

%%%%%%%%%%%%%%%%%%%
\begin{figure}[t]
\centering
\includegraphics[width=1.9in]{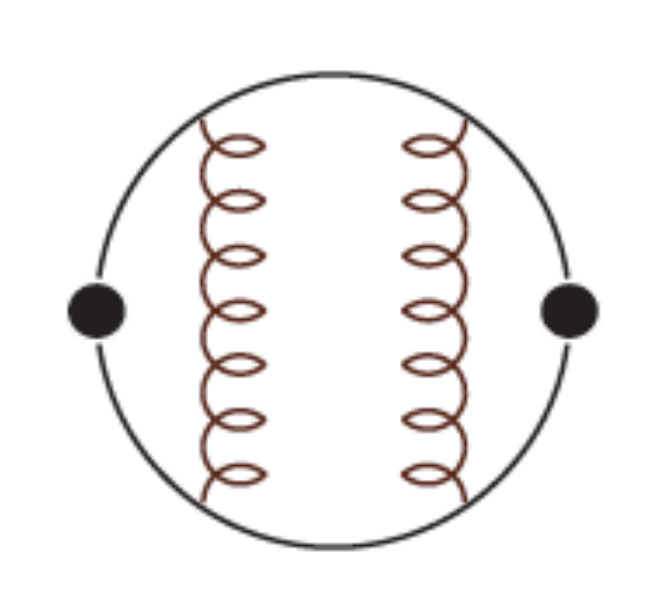}
\hspace{.1cm}
\includegraphics[width=3.7in]{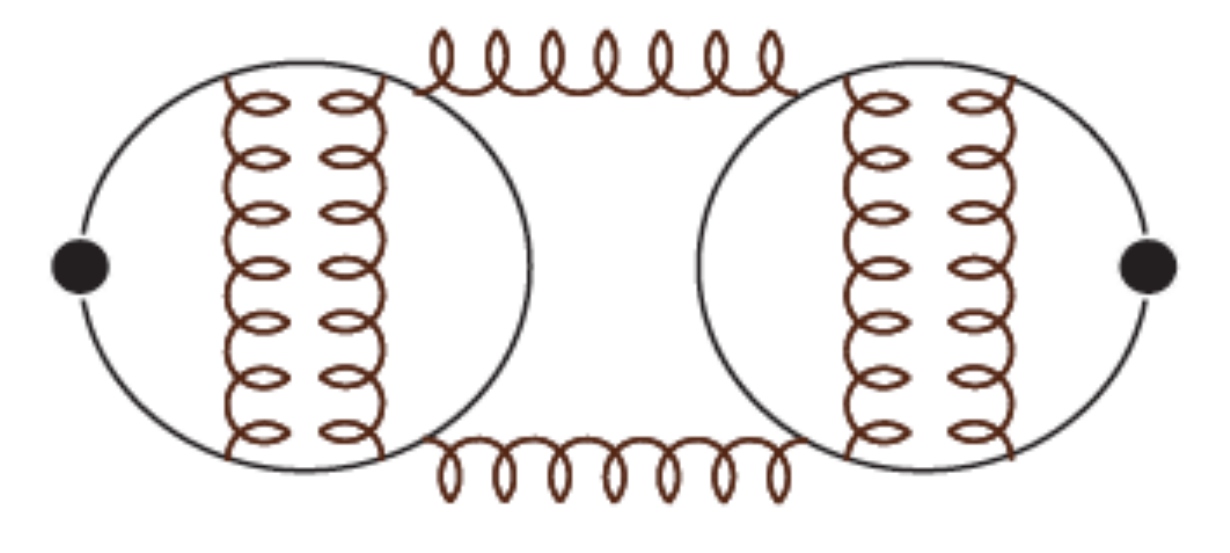}
\floatcaption{condiscon}{Examples of connected (left panel) and
disconnected (right panel) contributions to the vacuum polarization.
The black dots represent insertions of the current.}
\vspace*{2ex}
\end{figure}
%%%%%%%%%%%%%%%%%%%

The twist angle can be
chosen differently for the two quark lines in the connected part of the vacuum polarization, resulting in a continuously variable momentum flowing through the diagram. (Clearly, this trick does not work for the disconnected part.
For examples of connected and disconnected diagrams in this context,
see Fig.~\ref{condiscon}.)
If this momentum is chosen to be of the form~(\ref{allowed}),
then allowing $\theta_\m$ to vary over the range between 0 and $2\p$
allows $p_\m$ to vary continuously between $2\p n_\m/L_\m$ and $2\p (n_\m+1)/L_\m$.   This momentum is realized if, for example, we choose the anti-quark line in the vacuum polarization
to satisfy periodic boundary conditions (\ie, Eq.~(\ref{twistedbc}) with $\theta_\m=0$ for all $\m$), and the quark line twisted boundary conditions with twist angles
$\theta_\m$.\footnote{If anti-periodic boundary conditions in the time direction are chosen for
both quarks, the vacuum polarization still satisfies periodic boundary conditions.   Only the relative twist between the quark and anti-quark lines
introduces a twist in the boundary conditions for the vacuum polarization
as well.}

This choice can be viewed as following from the use of a mixed action 
\cite{DJ}.
The vacuum polarization is made out of two different valence quarks: one with periodic boundary conditions,
and one with twisted boundary conditions, but otherwise equal to the periodic
valence quark.   If the dynamical (sea) quarks are also periodic, the first valence quark is identical to the sea quark, but the twisted quark is not present in the sea, \ie, it is quenched.   In a path-integral definition of the theory, one would
thus introduce a ghost quark with the same twisted boundary conditions in order to cancel the determinant for the twisted quark.   Denoting the twisted
valence quark as $q_t$, as in Eq.~(\ref{twistedbc}), and the periodic
quark as $q$, the connected vacuum polarization then is a linear combination of terms of
the form\footnote{It is straightforward
to generalize our analysis to the choice of arbitrary twist angles in both
valence quarks.}
\begin{equation}
\label{JJ}
\langle J^+_\m(x)J^-_\n(y)\rangle=-\left\langle\tr\g_\m S_{q_t}(x,y)\g_\n S_q(y,x)\right\rangle\ ,
\end{equation}
with
\begin{eqnarray}
\label{twcurrent}
J^+_\m(x)&=&\qbar(x)\g_\m q_t(x)\ ,\\
J^-_\m(x)&=&\qbar_t(x)\g_\m q(x)\ ,\nonumber
\end{eqnarray}
and
where the trace is over Dirac and color indices. $S_q(x,y)$ is the full propagator
for the periodic quark $q$ and $S_{q_t}(x,y)$ is the full propagator for the twisted quark
$q_t$, equal to $\mbox{exp}(i\sum_\mu\theta_\mu(x_\mu-y_\mu)/L_\mu)$ times a periodic function of
$x$ and $y$ with period $L_\m$ in the $\m$ direction.   (In a slight abuse of 
notation, the average on the right-hand side of Eq.~(\ref{JJ}) is only over the gauge fields, while the average on the left is over both gauge and quark fields.)
The 
dependence of Eq.~(\ref{JJ}) on the twist angles $\theta_\mu$ is a finite-volume effect, and goes away in the limit $L_\m\to\infty$, in which all momenta become continuous.

The use of twisted boundary conditions immediately carries over to the lattice,
where, of course, we need to specify a discretization of the quark action.
In the following we will choose to use naive lattice quarks, but the discussion
generalizes to any choice of lattice quarks for which a conserved vector current can be defined.   In particular, our discussion applies directly to
staggered quarks as well.   The reason is that staggered quarks are nothing
else than naive quarks in a basis on which the gamma matrices are
diagonal, but with the resulting four-fold taste degeneracy removed.   This
diagonalization does not affect the discussion of the currents~(\ref{pscurrent})
below:   All one needs to do is replace the gamma matrices $\g_\m$ by the
staggered phases $\eta_\m(x)$, and drop the spin index on the quark fields.

For naive quarks with a nearest-neighbor Dirac operator, the currents~(\ref{twcurrent}) get replaced by the point-split currents
\begin{eqnarray}
\label{pscurrent}
j^+_\m(x)&=&\half\left(\qbar(x)\g_\m U_\m(x)q_t(x+\hmu)+
\qbar(x+\hmu)\g_\m U^\dagger_\m(x)q_t(x)\right)\ ,\\
j^-_\m(x)&=&\half\left(\qbar_t(x)\g_\mu U_\m(x)q(x+\hmu)+
\qbar_t(x+\hmu)\g_\m U^\dagger_\m(x)q(x)\right)\ ,\nonumber
\end{eqnarray}
where $U_\m(x)$ are the color gauge-field link variables.   

   In 
infinite volume, with conserved currents $j^\pm_\m(x)$, the construction of
a transverse vacuum polarization $\P_{\m\n}(x-y)$ on the lattice is then standard, and its
Fourier transform $\P_{\m\n}(p)$ takes the form~(\ref{Pem}) because of current conservation,\footnote{We are ignoring order-$a^2$ terms  of the form $\d_{\m\n}\sum_\k p^4_\k-p_\m^3 p_\n$, $(\d_{\m\n}p^2-p_\m p_\n)p^2_\n$, \etc\ 
For more discussion of Lorentz-covariance violating terms, see for example
Ref.~\cite{JLQCD}.}
 and one obtains $\P(p^2)$ by dividing by $p^2\d_{\m\n}-p_\m p_\n$ (for non-zero $p$).   However,
in finite volume, the boundary conditions break the symmetry that relates
$q$ and $q_t$, and thus the currents $j^\pm_\m(x)$ are not conserved for
non-vanishing $\theta_\mu$.   We derive the
corresponding modification of the relevant Ward--Takahashi identity in the next section, and discuss the construction of $\P_{\m\n}(x-y)$ in the
presence of twisted boundary conditions.

%%####%%
%\newpage
\section{\label{WI} Ward--Takahashi identity}
%%####%%
We consider the field transformations
\begin{eqnarray}
\label{fieldtr}
\d q(x)&=&i\a^+(x)e^{-i\theta x/L}q_t(x)\ ,\qquad \d\qbar(x)=-i\a^-(x)e^{i\theta x/L}\qbar_t(x)\ ,\\
\d q_t(x)&=&i\a^-(x)e^{i\theta x/L}q(x)\ ,\qquad\ \, \d\qbar_t(x)=-i\a^+(x)e^{-i\theta x/L}\qbar(x)\ ,\nonumber
\end{eqnarray}
in which we abbreviate
\begin{equation}
\label{abb}
\theta x/L = \sum_\m\theta_\m x_\m/L_\m\ ,
\end{equation}
and where $\a^\pm(x)$ are periodic functions of $x$.  The phases 
$\mbox{exp}(\pm i\theta x/L)$ have been inserted in order to ensure
that the transformed quark fields obey the same boundary conditions as
the untransformed fields.

Following the standard procedure, this transformation leads to the WTI
\begin{eqnarray}
\label{WTI}
&&\sum_\m\partial_\m^-\left\langle j^+_\mu(x)j^-_\nu(y)\right\rangle+
\frac{1}{2}\,\delta(x-y)\left\langle\qbar_t(y+\hnu)\gamma_\nu
U^\dagger_\nu(y)q_t(y)-\qbar(y)\gamma_\nu U_\nu(y)q(y+\hnu)\right\rangle
\nonumber\\
&&\hspace{0.7cm}-\frac{1}{2}\,\delta(x-\hnu-y)\left\langle
\qbar(y+\hnu)\gamma_\nu
U^\dagger_\nu(y) q(y)-
\qbar_t(y)\gamma_\nu U_\nu(y)q_t(y+\hnu)\right\rangle=0\ ,
\end{eqnarray}
where $\partial_\m^-$ is the backward lattice derivative, which, in Eq.~(\ref{WTI})
as well as Eqs.~(\ref{WTIsymm}) and~(\ref{WTIbr}) below always acts on $x$:
\begin{equation}
\label{bd}
\partial_\m^-f(x)=f(x)-f(x-\hmu)\ .
\end{equation}
If we would take $\theta_\m=0$ in all directions, the fields $q$ and $q_t$ would be identical, and this
identity would simplify to
\begin{eqnarray}
\label{WTIsymm}
&&\sum_\m\partial_\m^-\Biggl(\left\langle j^+_\mu(x)j^-_\nu(y)\right\rangle\\
&&+\frac{1}{2}\,\d_{\m\n}\d(x-y)\left\langle\qbar(y+\hnu)\gamma_\nu
U^\dagger_\nu(y)q(y)-\qbar(y)\gamma_\nu U_\nu(y)q(y+\hnu)\right\rangle\Biggr)=0\ .\nonumber
\end{eqnarray}
The quantity inside parentheses then defines $\P_{\m\n}(x-y)$, and its
Fourier transform is defined by
\begin{equation}
\label{FT}
\P_{\m\n}(x-y)=\frac{1}{V}\sum_p\,e^{ip(x-y)+i(p_\m-p_\n)/2}\,\P_{\m\n}(p)\ ,
\end{equation}
in which $V=\prod_\m L_\m$ and $p$ is summed over the momenta~(\ref{allowed}) with $\theta_\m=0$.   In momentum space, the WTI~(\ref{WTIsymm})
then takes the form
\begin{equation}
\label{WTImomentum}
\sum_\m \hp_\m\P_{\m\n}(p)=0\ ,\qquad
\hp_\m\equiv 2\sin{(p_\m/2)}\ ,
\end{equation}
from which the transverse form as in Eq.~(\ref{Pem}) follows (in the continuum limit).   The necessary presence of the contact term in Eq.~(\ref{WTIsymm}) is standard on the lattice \cite{KS}. 

With twisted boundary conditions, a natural generalization of Eq.~(\ref{WTIsymm}) is to define $\P_{\m\n}(x-y)$ similarly, but averaging
the contact term on the left-hand side of Eq.~(\ref{WTIsymm}) over the two
quark fields $q$ and $q_t$, leading to the definition
\begin{eqnarray}
\label{Pimunutw}
\P^{+-}_{\m\n}(x-y)&=&\left\langle j^+_\mu(x)j^-_\nu(y)\right\rangle\\
&&-\frac{1}{4}\d_{\m\n}\d(x-y)\Bigl(\left\langle\qbar(y)\gamma_\nu U_\nu(y)q(y+\hnu)
-\qbar(y+\hnu)\gamma_\nu
U^\dagger_\nu(y)q(y)\right\rangle
\nonumber\\
&&\phantom{\frac{1}{4}\d_{\m\n}\,\d(x-y)}+\left\langle\qbar_t(y)\gamma_\nu U_\nu(y)q_t(y+\hnu)-\qbar_t(y+\hnu)\gamma_\nu
U^\dagger_\nu(y)q_t(y)\right\rangle
\Bigr)\ .\nonumber
\end{eqnarray}
However, $\P^{+-}_{\m\n}(x-y)$ is not transverse, but instead obeys the identity
\begin{equation}
\label{WTIbr}
\sum_\m\partial_\m^-\P^{+-}_{\m\n}(x-y)+
\frac{1}{4}\left(\d(x-y)+\d(x-\hnu-y)\right)\langle j^t_\n(y)-j_\n(y)\rangle=0\ ,\\
\end{equation}
in which $j_\n(x)$ and $j_\n^t(x)$ are currents defined by
\begin{eqnarray}
\label{othercurrents}
j_\m(x)&=&\half\left(\qbar(x)\g_\m U_\m(x)q(x+\hmu)+\qbar(x+\hmu)\g_\m U^\dagger_\m(x)q(x)\right)\ ,\\
j^t_\m(x)&=&\half\left(\qbar_t(x)\g_\m U_\m(x)q_t(x+\hmu)+\qbar_t(x+\hmu)\g_\m U^\dagger_\m(x)q_t(x)\right)\ .\nonumber
\end{eqnarray}
It is important
to note that other choices for $\P^{+-}_{\m\n}(x-y)$ are possible, but there
will always be a non-vanishing contact term in the WTI.  The reason is that
the contact term in Eq.~(\ref{WTIbr}) (or, equivalently, in Eq.~(\ref{WTI})) cannot be written as a derivative, because
the fact that $q$ and $q_t$ fields satisfy different boundary conditions breaks
explicitly the isospin-like symmetry that otherwise would exist.  (For $\a^\pm$
constant and $\theta=0$, Eq.~(\ref{fieldtr}) is an isospin-like symmetry of the action.   As a check, we see
that for $q_t=q$, \ie, for $\theta=0$, the contact term
in Eq.~(\ref{WTIbr}) vanishes.)
The resulting non-transverse part of $\P^{+-}_{\m\n}$ therefore will need to be subtracted.      We discuss the properties of the contact term,
as well as its subtraction, in the next section.

For completeness, we also give the corresponding WTI for the case that
the local current
\begin{equation}
\label{localc}
j_\nu(y)=\qbar_t(y)\gamma_\nu q(y)
\end{equation}
is used instead of the current $j^-_\nu$ in the construction of the vacuum
polarization.   In that case, the WTI reads
\begin{equation}
\label{WTIlocal}
\sum_\mu\partial^-_\mu\left\langle j^+_\mu(x)j_\nu(y)\right\rangle+
\delta(x-y)\left\langle\qbar_t(y)\gamma_\nu q_t(y)-\qbar(y)\gamma_\nu q(y)\right\rangle=0\ .
\end{equation}
The structure of this WTI is the same as that of Eq.~(\ref{WTI}):  one obtains 
Eq.~(\ref{WTIlocal}) from Eq.~(\ref{WTI}) by omitting
the link variables in the contact term, and omitting $\nu$ in the arguments
of the fields and the delta function in Eq.~(\ref{WTI}).   Again, no vacuum
polarization can be constructed that is purely transverse.

%%####%%
%\newpage
\section{\label{subtraction} Subtraction of contact term}
%%####%%
Because of axis-reversal symmetry $\langle j_\n(y)\rangle=0$ in Eq.~(\ref{WTIbr}),\footnote{Recall that we assume dynamical quarks to have periodic, and
not twisted, boundary conditions.} but this is not true for $\langle j^t_\n(y)\rangle$, because twisted boundary conditions break this symmetry.
Instead, we have that
\begin{eqnarray}
\label{jtvev}
\langle j^t_\n(y)\rangle&=& -i\,\frac{c}{a^2}\,\hth_\n\left(1+O(
\hth^2)\right)\ ,\\
\hth_\m&=&\theta_\m/L_\m\ ,\nonumber
\end{eqnarray}
where $c$ is a numerical constant, and
where we made the lattice spacing $a$ explicit.   This expansion is valid
when $\hth$ is small compared to both $1/a$ and the quark mass $m$.
Equation~(\ref{jtvev}) follows from 
dimensional analysis and the
fact that if we let $\theta_\m\to-\theta_\m$
under an axis reversal in the $\m$ direction, this axis reversal would be a symmetry of the
theory.  The contact
term in Eq.~(\ref{WTIbr}) is quadratically divergent (at fixed $L_\m$), and cannot be ignored.

The vacuum polarization $\P^{+-}_{\m\n}(x-y)$ defined in
Eq.~(\ref{Pimunutw}) can be written
as
\begin{equation}
\label{pervacpol}
\P^{+-}_{\m\n}(x-y)=e^{ia\theta(x-y)/L}\,F^{+-}_{\m\n}(x-y)\ ,
\end{equation}
where  
$F^{+-}_{\m\n}(x-y)$ is a periodic function of $x-y$ with period $L_\m$ in
the $\m$ direction.   This implies that the Fourier transform of $\P^{+-}_{\m\n}(x-y)$ is defined as in Eq.~(\ref{FT}), but now with the momentum
$p$ summed over the values~(\ref{allowed}).  Let us decompose\begin{equation}
\label{vacpolpar}
\P^{+-}_{\m\n}(\hp)=\left(\hp^2\d_{\m\n}-\hp_\m\hp_\n\right)\P^{+-}(\hp^2)+
\frac{\d_{\m\n}}{a^2}\,X_\n(\hp)\ ,
\end{equation}
in which a quadratically divergent term $X_\n(\hp)$ has been
added to the transverse part, in order to accommodate the explicit breaking
term in the WTI~(\ref{WTIbr}).  In momentum space,
Eq.~(\ref{WTIbr}) takes the form
\begin{eqnarray}
\label{WTIms}
i\sum_\m\hp_\m\P^{+-}_{\m\n}(\hp)&=&
-\cos{(ap_\n/2)}\langle j^t_\n(0)\rangle\\
&=&\frac{i}{a^2}\,\hp_\n X_\n(\hp)\ ,\nonumber
\end{eqnarray}
where in the second line we substituted Eq.~(\ref{vacpolpar}).
In the appendix, we verify Eqs.~(\ref{WTIms}) and~(\ref{jtvev}) to one loop.
Using Eq.~(\ref{jtvev}), we find
\begin{eqnarray}
\label{Xsol}
X_\n(\hp)&=&\frac{i}{2}\cot{\left(ap_\n/2\right)}\,a^3\langle j^t_\n(0)\rangle\\
&=&
\half\, c\,\cot{\left(ap_\n/2\right)}\,a\hth_\n
\left(1+O(
\hth^2)\right)\ .\nonumber
\end{eqnarray}
Note that this result for $X_\n(\hp)$ does {\it not} have a 
pole.   The cotangent only has a pole when its argument is equal to zero,
modulo $\p$.   But, using Eq.~(\ref{allowed}), this would require that
\begin{equation}
\label{nopole}
\p n_\n+\theta_\n/2=k\p L_\n/a\ ,\qquad k\ \mbox{integer}\ ,
\end{equation}
which is only possible for $\theta_\n=0$, given the allowed
range for $\theta_\n$.   However, for $\theta_\n=0$, $\langle j^t_\n(0)\rangle=0$, and thus $X_\n=0$ as well.  Equations~(\ref{WTIms}) 
and (\ref{Xsol}) show
how to compute $X_\n(\hp)$ from $\P^{+-}_{\m\n}(p)$
or $\langle j^t_\n(0)\rangle$, while Eq.~(\ref{Xsol})
makes the momentum dependence explicit.   One can then
subtract $\d_{\m\n}X_\n(\hp)/a^2$ from $\P^{+-}_{\m\n}(p)$,
so that $\P^{+-}(\hp)$, needed in order to compute $a_\m^{\rm HLO}$
(\seef\ Eq.~(\ref{amu})), can be determined.  Of course, for $\m\ne\n$ 
no subtraction is needed.

Some comments are in order.   First
we wish to emphasize that the counter term proportional to $X_\n$ is a pure
finite-volume artifact; it will disappear in the 
infinite-volume limit at fixed lattice spacing.   If we keep $p_\n=(2\p n_\n+\theta_\n)/L_\n$ fixed 
in the factor $\cot(ap_\n/2)$ in Eq.~(\ref{Xsol}) while taking $L_\n\to\infty$, clearly 
$X_\n$ goes to zero, and thus
there is no need for a counter term in Eq.~(\ref{vacpolpar}). 

Then, the decomposition~(\ref{vacpolpar}) is not
exact at non-zero lattice spacing in a finite volume.   The only tensor structure
allowing for a power-like divergence with the lattice spacing is $\d_{\m\n}$,
as in Eq.~(\ref{vacpolpar}), because $\d_{\m\n}$ is the only $2$-index tensor with mass
dimension zero.  But, for example, there
could be terms proportional to $\hth_\m\hth_\n$, $\hp_\m\hth_\n$, $\hth_\m\hp_\n$, or $\hp_\m\hp_\n$ that are not transverse.   Such terms are not divergent in the continuum
limit, because both $\hp_\m$ and $\hth_\m$ have mass dimension one.
Furthermore, since any non-transverse terms originate
from the use of twisted boundary conditions, such terms must have at least one
factor of $\hth_\m=\theta_\m/L_\m$, \seef\ Eq.~(\ref{jtvev}).   Therefore, they disappear in the 
infinite-volume limit.\footnote{Other tensor structures are possible on the lattice, but these are all of order $a^2$, and thus vanish in the continuum
limit, see footnote 6.}

%%####%%
%\newpage
\section{\label{numerical} Numerical tests}
%%####%%
\begin{figure}[t]
\centering
\includegraphics[width=6in]{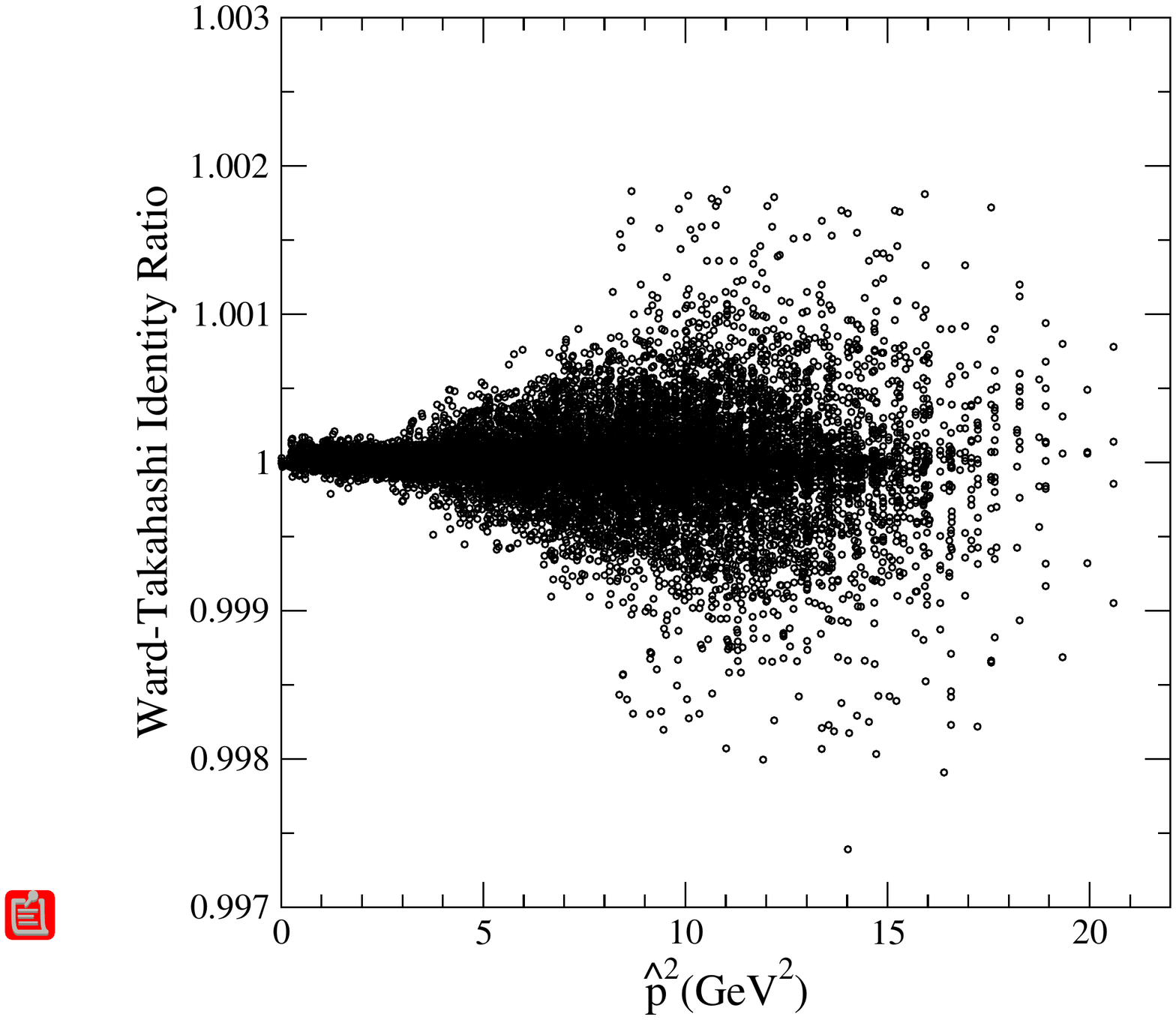}
\floatcaption{f2}{Ratio of the left-hand and right-hand sides of the WTI~(\ref{WTIms}), for a typical gauge-field configuration from the
asqtad MILC ensemble with $L^3\times T=48^3\times 144$, $1/a=3.35$~GeV,
$am=0.0036$, $\theta_x=\theta_y=\theta_z=0.28\pi$, $\theta_t=0$.}
\vspace*{2ex}
\end{figure}
The WTI~(\ref{WTIms}) holds on each gauge configuration, making
it straightforward to test the identity numerically.   We did so on a number
of configurations from a MILC asqtad ensemble, always finding the two
sides of the WTI to agree within the numerical precision employed.

As an example, in Fig.~\ref{f2} we show the ratio of both sides of the WTI~(\ref{WTIms}),
computed on a typical configuration (for parameters, see the
figure caption).   The conjugate-gradient stopping condition was $10^{-8}$
on the residual.      We observe that the WTI is satisfied with a 
precision of order a few permille, and about an order of magnitude better than
that for $\hp^2\,\ltap\,4$~GeV$^2$.
We replaced $\langle j^t_\n(0)\rangle$ on the right-hand side of 
Eq.~(\ref{WTIms}) by $\langle j^t_\n(0)-j_\n(0)\rangle$, conform Eq.~(\ref{WTIbr})
(for the arbitrary choice $y=0$),
because on a single configuration generically
$\langle j_\n(0)\rangle\ne 0$.\footnote{In addition, for the same reason,
$\P^{+-}_{\m\n}$ is not translation invariant, and thus a function of the
source and sink points $x$ and $y$.}

\begin{figure}[t]
\centering
\includegraphics[width=6in]{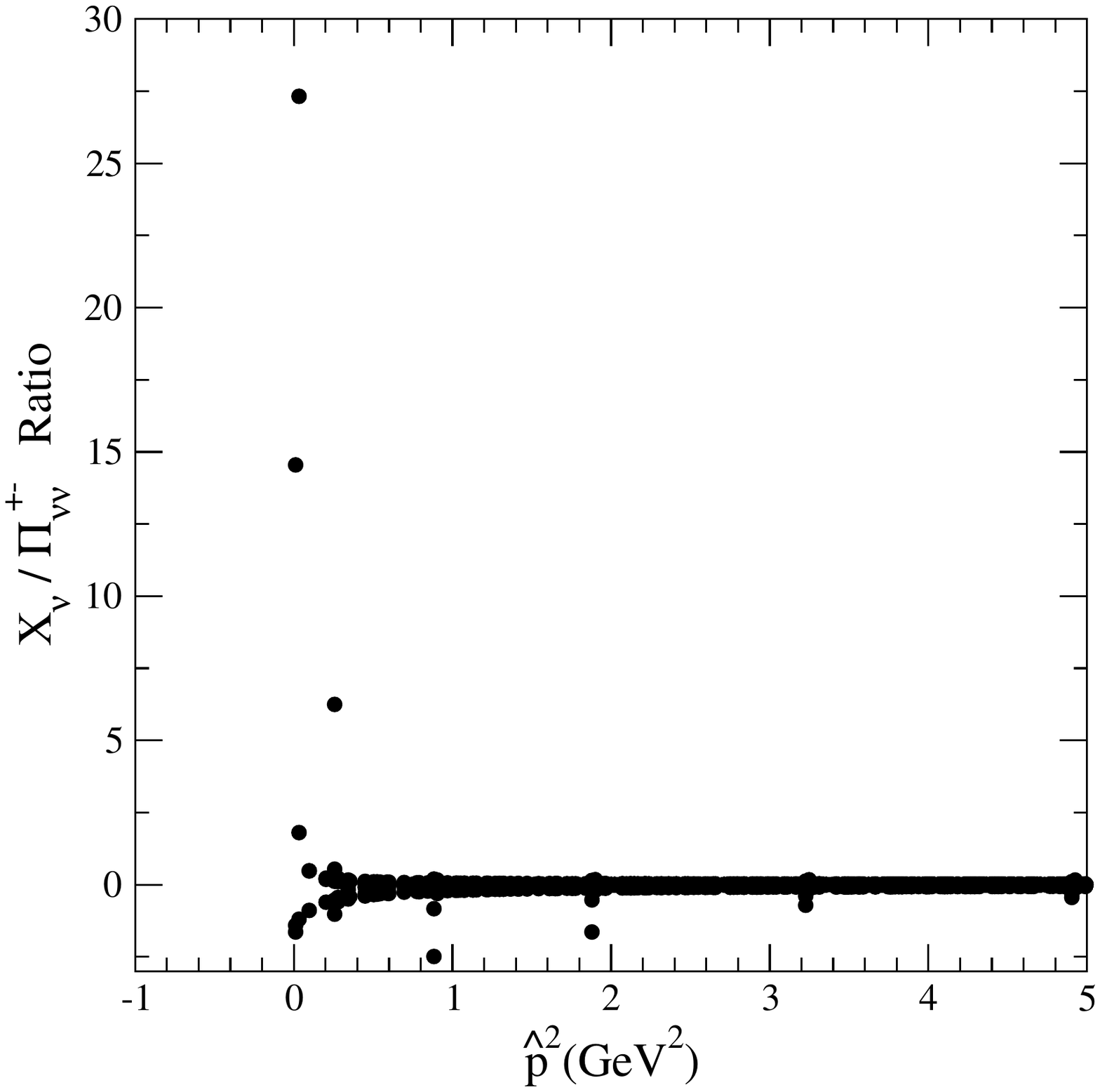}
\floatcaption{f3}{Ratio of the second term on the right-hand side and the 
left-hand side of Eq.~(\ref{vacpolpar}), for the same gauge-field configuration
as used for Fig.~\ref{f2}.}
\vspace*{2ex}
\end{figure}

In Fig.~\ref{f3} we show, for the same gauge field configuration, the ratio
\begin{equation}
\label{ratio}
\frac{X_\n(\hp)}{a^2\P^{+-}_{\n\n}(\hp)}\ ,
\end{equation}
with choices for the momentum $\hp$ such that the denominator does not 
vanish.  Here $\P^{+-}_{\m\n}(\hp)$ was obtained as the Fourier transform
of $\P^{+-}_{\m\n}(x)$, taking $y=0$ in Eq.~(\ref{Pimunutw}),
and $X_\n(\hp)$ was obtained from Eq.~(\ref{Xsol}), with again
$\langle j^t_\n(0)\rangle$ replaced by $\langle j^t_\n(0)-j_\n(0)\rangle$.

In Fig.~\ref{f3} we see that for some momenta (especially in the low-momentum region) the size of the counter term in Eq.~(\ref{vacpolpar})
can be significant.   We also find that averaging (over volume or over
configurations) appears to reduce the effect of the counter term.
While the effect of averaging is at present still under investigation, the result shown in the
figure indicates that at least on single configurations the effect of the
counter term cannot be ignored.

%%####%%
%\newpage
\section{\label{conclusion} Conclusion}
%%####%%
In this article, we investigated the use of twisted boundary conditions in a finite volume, in order to compute the hadronic vacuum polarization on a lattice for all values of the euclidean momenta, instead of only those allowed with periodic
boundary conditions.    As we explained in the Introduction, this is important for a high-precision computation of the leading-order hadronic contribution to the
muon anomalous magnetic moment.

In order to vary the momentum flowing through the vacuum polarization, the quark and anti-quark lines constituting the connected contribution to the vacuum polarization should obey boundary conditions with different twist
angles.\footnote{As mentioned before, this method therefore does not apply to the disconnected part.}   This implies that the isospin-like symmetry relating these two 
quark lines is broken explicitly.   This breaking shows up as an extra contact term in the relevant Ward--Takahashi identity that cannot be removed by a local redefinition of the currents, and which we 
showed to be quadratically divergent.   Correspondingly, the vacuum polarization is not transverse, but instead contains a quadratically divergent
term which needs to be subtracted.   We emphasize that this extra term is a 
finite-volume artifact caused by the use of twisted boundary conditions.
A consequence of this is that the point-split currents considered in this
article still do not renormalize, so that no $Z$ factors appear if the currents
$j^\pm_\mu$ of Eq.~(\ref{pscurrent}) are used in order to define the vacuum polarization.\footnote{As usual, a non-trivial $Z$ factor appears if the local
current $j_\mu$ of Eq.~(\ref{localc}) is used.}

The analysis leading to this conclusion also leads to a recipe for removing the unwanted term from the vacuum polarization, as discussed in Sec.~\ref{subtraction}.   The subtracted vacuum polarization is still not exactly
of the desired form~(\ref{Pem}), but the remaining violations are lattice
and finite-volume artifacts that should automatically disappear in the continuum and infinite-volume limits.   Without the removal of the quadratically divergent term this would not the case.  Of course, while the quadratically
divergent nature of the contact term tends to increase the importance of this effect
at smaller lattice spacing, the fact that it is a finite-volume effect will help
suppress the effect on larger volumes.
It thus remains to
be seen how numerically significant the effect is in practice on realistic lattices with a given lattice spacing.   

\vspace{3ex}
%\newpage
\noindent {\bf Acknowledgments}
\vspace{3ex}

We would like to thank USQCD for the computing resources used to generate the vacuum polarization as well as the MILC collaboration for providing the configurations used.
TB and MG are supported in part by the US Department of Energy under
Grant No. DE-FG02-92ER40716 and Grant No. DE-FG03-92ER40711.
MG has also been supported in part by the Spanish Ministerio de Educaci\'on, Cultura y Deporte, under program SAB2011-0074.
SP is supported by CICYTFEDER-FPA2011-25948, SGR2009-894,
the Spanish Consolider-Ingenio 2010 Program
CPAN (CSD2007-00042).

%%####%%
%\newpage
\appendix
\section{\label{one-loop} Vacuum polarization at one loop}
%%####%%
In this Appendix, we verify the WTI and the occurence of the quadratic
divergence in $\P_{\m\n}(x-y)$ at one loop.   We set $a=1$ again.
At one loop, the vacuum polarization of Eq.~(\ref{Pimunutw}) is just that in the
theory of free quarks, and, using the Feynman rules for naive fermions,
we find
\begin{eqnarray}
\label{vacpoloneloop}
\P^{+-}_{\m\n}(p)&=&-\frac{N_c}{V}\sum_k\tr\left[\g_\m\,
\frac{\cos\left(k_\m+p_\m/2\right)}{i\sum_\k\g_\k\sin(k_\k+p_\k)+m}
\,\g_\n\,
\frac{\cos\left(k_\n+p_\n/2\right)}{i\sum_\l\g_\l\sin{k_\k}+m}\right]\\
&&\hspace{-0.5cm}+\frac{i}{2}\,\d_{\m\n}\,\frac{N_c}{V}\sum_k\tr\left[\g_\n\left(
\frac{\sin{k_\n}}{i\sum_\k\g_\k\sin{k_\k}+m}
+\frac{\sin{(k_\n+\hth_\n)}}{i\sum_\k\g_\k\sin{(k_\k+\hth_\k)}+m}\right)
\right]\ ,\nonumber
\end{eqnarray}
in which $p$ is one of the momenta specified in Eq.~(\ref{allowed}), and $k$ 
is summed over periodic momenta, with components $k_\m=2\p n_\m/L_\m$
with $n_\m\in\{0,\dots,L_\m-1\}$; $N_c$ is the number of colors.

Using trigonometric identities, it is straightforward to show that
\begin{eqnarray}
\label{WTIoneloop}
i\sum_\m\hp_\m\P^{+-}_{\m\n}(p)&=&\\
&&\hspace{-2.5cm}
-2i\cos{(p_\n/2)}\frac{N_c}{V}\sum_k
\left(\frac{\sin(2k_\n)}{\sum_\k\sin^2{k_\k}+m^2}
-\frac{\sin(2(k_\n+\hth_\n))}{\sum_\k\sin^2{(k_\k+\hth_k)}+m^2}\right)
\nonumber\\
&&\hspace{-3cm}=\ 2i\cos{(p_\n/2)}\,\hth\left[\frac{N_c}{V}\sum_k
\left(\frac{2\cos(2k_\n)}{\sum_k\sin^2 k_\k+m^2}
-\frac{\sin^2(2k_\n)}{(\sum_k\sin^2 k_\k+m^2)^2}\right)\right]+O(\hth^3)\ .
\nonumber
\end{eqnarray}
The quantity in square brackets on the last line is quadratically
divergent in the continuum limit, and leads to a one-loop result for $\langle j^t_\n(y)\rangle$
of the general form~(\ref{jtvev}).
In manipulating the sums over $k$, one should keep in mind that only
shifts by periodic momenta are allowed.   For instance, to obtain Eq.~(\ref{WTIoneloop}), we made use of the shift $k_\m\to k_\m-p_\m+\hth_\m$,
which is allowed because $p_\m-\hth_\m=2\p n_\m/L_\m$ for some
integer values of $n_\m$.   A shift $k_\m\to k_\m-\hth_\m$ is {\it not}
allowed, and thus the right-hand side of Eq.~(\ref{WTIoneloop}) does not
vanish for $\hth_\n\ne 0$.   Equations~(\ref{vacpoloneloop}) and
(\ref{WTIoneloop}) hold also for staggered fermions, if the right-hand side
of these two equations is divided by four.

  %%%%%%%%%%%%%%%%%%%%%%%%%%%%%%%%%%%%%%%%%%%%%%%%%%%%%%%%%%%%%%%
%\newpage

\end{document}